\documentclass[conference, 10pt]{IEEEtran}
\pdfoutput=1
\IEEEoverridecommandlockouts
\usepackage{cite}
\usepackage{amsmath,amssymb,amsfonts}
\usepackage{algorithmic}
\usepackage{graphicx}
\usepackage{textcomp}
\usepackage{xcolor}
\def\BibTeX{{\rm B\kern-.05em{\sc i\kern-.025em b}\kern-.08em
    T\kern-.1667em\lower.7ex\hbox{E}\kern-.125emX}}
\begin{document}


\title{How to integrate with real cars - minimizing lead time at Volkswagen}

\author{\IEEEauthorblockN{1\textsuperscript{st} Jan Kantert}
\IEEEauthorblockA{\textit{Digital Services \& Data Analytics} \\
\textit{Volkswagen Commercial Vehicles}\\
Hanover, Germany \\
jan.kantert@volkswagen.de}
\and
\IEEEauthorblockN{2\textsuperscript{nd} Michael Nolting}
\IEEEauthorblockA{\textit{Digital Services \& Data Analytics} \\
\textit{Volkswagen Commercial Vehicles}\\
Hanover, Germany \\
michael.nolting@volkswagen.de}
}

\IEEEpubid{ICSE-SEIP-38}

\maketitle

\begin{abstract}
The most successful tech companies of the world release new software versions to production multiple times a day.
Thereby, they are able to quickly fix emerging bugs and rapidly deliver new features to their customers.
This leads to short development cycles, minimal lead times and a high customer-centricity.
Short development cycles are easy to achieve if you start a software project on a green field.
Nevertheless, this does not apply to brown field environments which are usually found in big corporates such as traditional car manufacturers.
For instance, if you want to integrate with real cars you have to interface legacy systems with development cycles of up to several months. 
We present a solution, which worked for one of the world's largest car manufacturer, leveraging in-house core development teams, dynamic stages and feature-toggles to overcome a brown field environment, allow for short development cycles and minimize the lead time.
\end{abstract}


\section{The problem: Minimizing lead time}

\begin{figure}[htbp]
\centerline{\includegraphics[width=\linewidth]{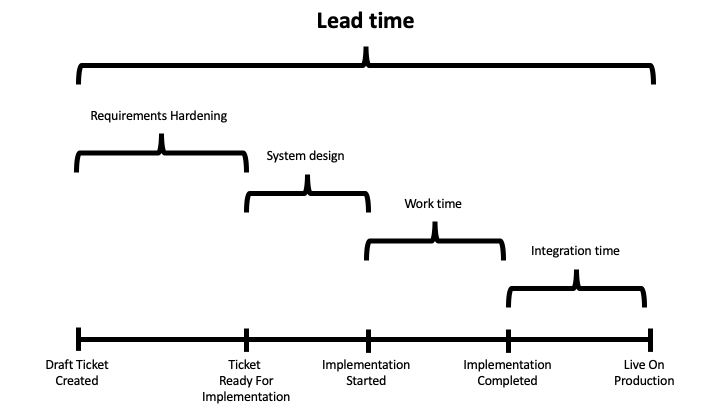}}
\caption{Lead time and its phases in a corporate environment}
\label{fig:lead-time}
\end{figure}

Today, tech companies such as Google, Amazon or Facebook adopt agile methods, DevOps principles and modern software tools to deploy changes to production hundreds or even thousands of times per day.
In an age where competitive advantage requires fast time to market and relentless experimentation, organizations that are unable to replicate these outcomes are cursed to lose in the marketplace to more adaptive competitors and may go out of business entirely, much like the manufacturing organizations that did not adopt Lean principles in the past\cite{devops}.


The lead time measures the performance in the technology value stream, i.e. how fast a company is able to deliver new software products or additional features to their customers.
Car manufacturers currently suffer from a lead time of several months regarding the development of digital services.
This is mainly driven by the fact, that their integration process requires a long time which is a common problem in large complex organizations that are working with tightly-coupled, monolithic applications, scarce integration test environments, long test and production environment lead times, high reliance on manual testing, and multiple required approval processes\cite{rethink-ee}.
Due to various safetly regulations multiple test and approval gates are implemented depending on the type of software\cite{iso26262}.
All components within the car (i.e. electronic control units) got very long development cycles while cloud backend with less regulation can iterate much faster.
When this occurs, our value stream may look like figure \ref{fig:lead-time}, where each phase may last weeks or months on its own.

\section{The Solution: Own your code! Own your product!}

To break free from legacy hell and minimize lead time we had to achieve the following goals at Volkswagen:

\IEEEpubidadjcol

\begin{enumerate}
	\item{In-house code hosting, build server and an in-house core development team}
	\item{Dynamic stages for fast feedback by early integration}
	\item{Minimizing risk and experimentation in the car approval process}
\end{enumerate} 

\subsection{In-house code hosting, build server and an in-house core development team}

Most IT development activity happened outside of Volkswagen in the 
past which had multiple adverse effects on the overall progress \cite{sw-dev-early-stage}.
A lot of internal teams lost the capabilities to understand how their systems work.
As a result, they were no longer able to estimate efforts for new features or to determine where failures were originating.
Operations became challenging.

To solve this dilemma, we started to build up an in-house core development team to re-own our own products technically.
Our internal mission was: Own your code! Own your product!
The idea was not to replace external developers but to meet them at eye level.
As a result, we could leverage feedback from our suppliers (as we had a common basis for discussion now) and build a better product.

Due to the previous outsourcing process and mentality, hosting of source code and the build systems for one project were often distributed over multiple suppliers.
That created barriers and also lead to very heterogeneous development environments.
Getting access became a nightmare and any efforts to perform large scale refactoring were hampered.
Building a single release often took days to weeks in such environments.

We solved this by building up a shared CI/CD chain and hosted the source code within the company.
That allowed us to share infrastructure between projects and lower the barrier for entrance.
We also noticed that suppliers started to cooperate much more as a result of this as they could now access all the code.
In addition to this, we conducted quaterly Scrum-based product increment plannings and relied on a scaled agile framework (such as SAFEScrum) to scale the development \cite{scale-agile} and harmonize the system's design process.

\subsection{Dynamic stages for fast feedback by early integration}

Automotive companies need to maintain their backends for approximately 10 to 15 years as part of the vehicle's life-cycle.
As a result, a lot of legacy systems exist for various car platforms (i.e. there is a new platform every few years and each of them is maintained for a long time).
To develop code in such an environment you need a lot of credentials for many systems which is almost impossible for an individual contributor to gather.
Consequently, integration of code usually happens rarely and very late in the development cycle.
As a matter of fact, postponed releases are the norm and not the exception.

To solve this challenge, we started to offer dynamic stages (similiar to Google's pre-submit infrastructures \cite{presubmit-google}) to developers which allow them to integrate early and often for fast feedback.
Technically, we spin up all microservices in the cloud and connect them to legacy systems (i.e. email, authentification providers, vehicle backends, vehicle master data, vehicle diagnostic error codes etc).
Developers can choose if they want mocked legacy systems or QA/approval instances.
If the corresponding system shares the infrastructure, we can even spin up the latest development version of the other project.
When using QA instances, it is possible to test drive the development stage with real vehicles (if the vehicle is configured for the approval backend).
This can happen in a pre-merged state so that code can be tested before hitting the development branch of our version control system.

For all those services, we try to offer maximum self-service for developers, testers and quality insurance experts.
Previously, they had to communicate with a lot of people just to get a single test stage.
Now we can have one stage per feature within a few minutes and test features early.
To minimize expose of credentials for legacy systems, we contain those in the cloud environment but keep them available to everybody.
For instance, we offer self-service tools to request tokens which are valid for a limited time but share certificates to create tokens.

\subsection{Minimizing risk and experimentation in the car approval process}

Eventually, our new software has to be approved as part of a car's start of production (SoP).
The entire SoP process usually takes five years and software should be completed at latest one year before the car is produced.
That certainly needs to be the case for all components within the vehicle but it is not realistic for backend components in the cloud which can be updated in a matter of minutes.
Before the SoP or generally on any release, however, the eletronic development and quality assurance departments will ensure that the software still works with our vehicles.
As you can imagine this is a very conservative process which is conducted only a few times a year and if you release rarely you will have large changes and experience much pain.

To reduce friction in this process, we heavily rely on feature flags \cite{feature-flags} and create deterministic builds with fixed version numbers over all micro services.
We need one version for the process to track bugs over all the involved parties.
Later we need to be able to know exactly which software commits were included as development goes on and the manual tests take a few days at least.
With feature flags we can fail easily if certain features lack maturity and do not block the process.
Even if all of our new features fail we should still be able to deploy the previous feature set.
It also allows us to include small experiments into a release early on and disable them if they cause issues.
Furthermore, we build trust with other parties in the approval process which enables us to push for more frequent releases even if they are pure bugfix releases.  

\section{Future work and next steps}
To improve this further we started automated integration tests and end to end tests.
This allows us to test integration between systems in a pre-merged state which is great feedback for developers.
Furthermore, we can automate certain tests with real vehicles by simulating the backend communication of vehicles.
As part of releases we run those tests as smoke tests to make sure that configuration and permissions of legacy systems are correct which has been a source of issues when releasing new features in the past.



\end{document}